\documentclass[preprint,12pt]{elsarticle}

\usepackage[utf8]{inputenc}
\usepackage{graphicx}
\usepackage{amssymb}

\usepackage{lineno}
\usepackage{color}

\usepackage{ulem}

\usepackage{cancel}

\long\def\/*#1*/{} 

\journal{Optics and Lasers in Engineering}

\begin{document}

\begin{frontmatter}


\title{Description of a multifocal arrangement of asymmetric Kummer-beam optical vortices.}



\author[udea]{Natalia Londoño}
  \author[udea]{Edgar Rueda}
  \author[poli]{Jorge A. Gómez}
\author[ciop]{Dafne Amaya}
\author[azul]{Alberto Lencina}

\address[udea]{Grupo de Óptica y Fotónica, Instituto de Física, Universidad de Antioquia U de A, Calle 70 No. 52-21, Medellín, Colombia}
\address[poli]{Grupo de Física Básica y Aplicada, Politécnico Colombiano Jaime Isaza Cadavid, Medellín, Colombia}
\address[ciop]{Centro de Investigaciones Ópticas, CONICET-UNLP-CIC, P.O. Box 3, 1897 Gonnet, Argentina}
\address[azul]{Laboratorio de Análisis de Suelos, Facultad de Agronomía, Universidad Nacional del Centro de la Provincia de Buenos Aires, CONICET, P.O. Box 47, 7300, Azul, Argentina}

\begin{abstract}
In this work, an analytic expression is derived for the optical field generated by an off-axis Gaussian beam  diffracted by a Discrete Vortex-Producing Lens. With this system, a multifocal arrangement of asymmetric optical vortices is obtained whose topological charge values change with the position along the optical axis. This scheme allows obtaining a principal asymmetric vortex corresponding with the topological charge value of the phase mask and secondary ones with different charges. With the analytical expression, the effects induced by the discretization and misalignment on the irradiance and phase of each vortex can be simultaneously studied. A signal-to-noise ratio expression is derived to verify if the noise of the multifocal system might affect importantly the optical vortices of interest. Finally, we proposed new metrics to measure the off-axis displacement using the phase of the vortex. We concluded that noise is not a problem, then a Discrete Vortex-Producing Lens can be used as a continuous Spiral Phase Plate with the bonus that secondary optical vortices can be used in the same way as principal optical vortices.

\end{abstract}


\end{frontmatter}


\section{Introduction}
\label{S:1}
Optical vortices (OVs) are localizations in space, lines in 3D or points in 2D, where the amplitude of the field is zero while its phase is undefined \cite{1,2}.Around them, the phase has a helical shape represented by the term $\exp(i\ell\theta)$, where $\theta$ is the azimuthal angle and $\ell$ is the so-called topological charge that represents the number of times the wavefront phase varies $2\pi$ \cite{3}. One important method used to generate optical vortices consists of a beam with a Gaussian profile impinging on a spiral phase mask with a linear dependence between 0 and $2\pi \ell$. Then, an OV with a doughnut-shape irradiance pattern, i.e. with radial symmetry, is obtained at the far-field. Such OVs can be easily produced from diffractive optical elements such as computer-generated holograms (CGHs) \cite{6}, spiral phase plates (SPPs) \cite{7} or vortex-producing lenses (VPLs) \cite{8,9}. Because of OVs' intrinsic features and relative easiness for experimental generation, interest in the study of these types of beams has grown, leading to some interesting applications in metrology \cite{10}, phase-shifting interferometry \cite{12}, stellar choronography \cite{13}, optical tweezers \cite{14,15}, optical communication \cite{16}, among others. In recent works, it has been shown both, analytically and numerically, that relative displacements between the input Gaussian beam and the phase mask generating the vortex, lead to loss of the irradiance radial symmetry \cite{17,18,19,20,21}. Particularly,  Anzolín et al. \cite{19} using a continuous spiral phase plate (CSPP) developed an analytical model to measure off-axis displacements by measuring the asymmetry of the irradiance pattern. They showed that OVs with higher topological charges tend to lose their irradiance symmetry faster than those of lower charges, meaning that higher topological charges are more sensitive to misalignments. This behavior was experimentally verified using binary VPLs \cite{22}. Binary or discretized phase masks differ from the one proposed by Anzolin et al. in that they produce multifocal arrangements with multiple OVs \cite{23,24,26,27}. In particular, Rumi et al. \cite{24} show analytically that an on-axis Gaussian beam impinging into a DVPL produces secondary optical vortices (SOVs) at different positions of the optical axis, depending on the topological charge of the principal vortex $\ell$ and the number of discretization levels $N$. Another important aspect about those works concerning off-axis beams is that they only study the irradiance behavior, and occasionally the angular momentum \cite{18,Bess,LG}, paying little attention to the phase of the optical field.

In this work, based on the analytical derivations of \cite{19,24}, we develop the analytical expression of the optical field for the general case of an off-axis Gaussian beam impinging on a discretized vortex-producing lens. The obtained field is expressed as a summation of Kummer-beams. By varying the setup parameters, off-axis displacement, and discretization levels, we analyze the effects over the irradiance and phase of the optical field at some observation plane. Also, we study the relevance of the noise of the multifocal system with respect to the OV of interest in that plane. Finally, the behavior of the intensity and phase distributions are studied as the misalignment is varied.

\section{Analytic description}
\label{S:2}

\begin{figure}[t]
\centering\includegraphics[width=8 cm]{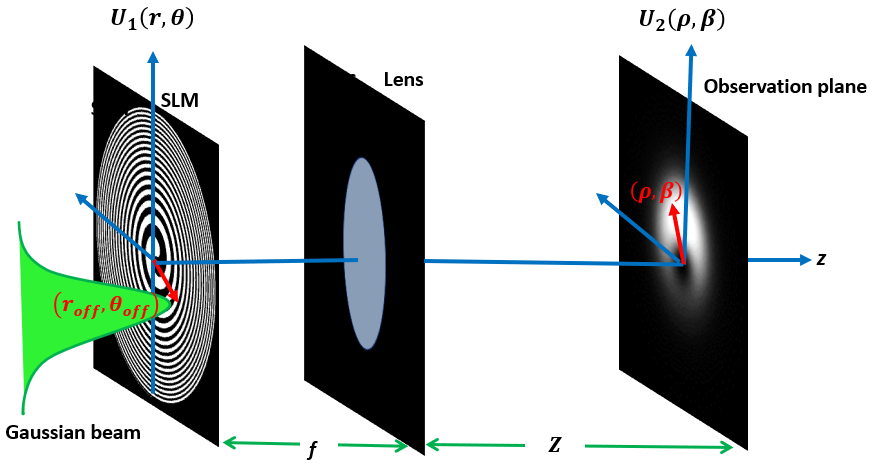}
\vspace{-0.3cm}
\caption{Geometric configuration for Fresnel propagation in free space of a Gaussian beam incident off-axis onto a DVPL.} \label{f:1}
\end{figure}

An off-axis Gaussian beam centered at $(r_{off},\theta_{off})$ (see Fig. \ref{f:1}), impinges normal to the surface of a DVPL located at plane SLM.

\begin{figure}[t]
\centering
\includegraphics[width=8 cm]{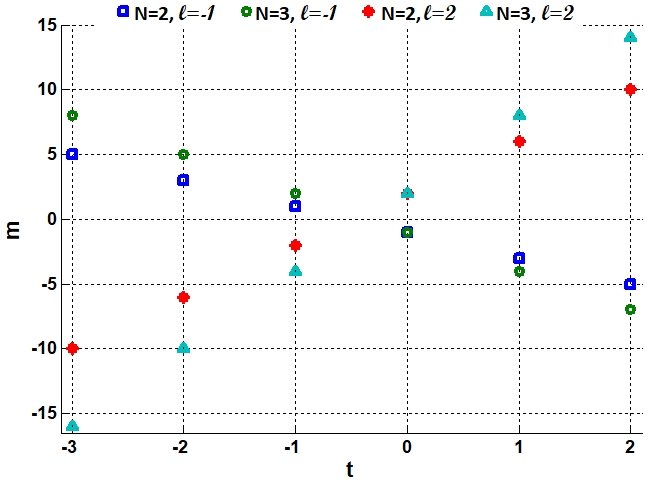}
\vspace{-0.3cm}
\caption{Secondary topological charge values $m$ as a function of $t$ for different values of $N$ and $\ell$.}\label{f:2}
\end{figure}

In polar coordinates, the transmission function of the DVPL can be expressed mathematically by \cite{24}:

\begin{equation}
\label{eq:01}
\exp\left(j\Phi(r,\theta)\right)=\exp\left(j\frac{2\pi}{N}\textup{floor}\left[\frac{N}{2\pi}\left(\ell\theta-\frac{kr^2}{2f_{FR}}\right)\right]\right),
\end{equation}

\noindent where $f_{FR}$ is the phase diffractive-lens focal distance, $k$ is the wavenumber, $N$ is the number of discretized phase levels, $\ell$ is the principal topological charge and $\textup{floor}[x]$ is a function taking the nearest integer smaller than or equal to $x$. This transmission function can be expanded in a Fourier series on a base of topological charges $m$:

\begin{equation}
\label{eq:02}
e^{j\Phi(r,\theta)}=\sum_{m=-\infty}^{\infty}T_m(r)e^{jm\theta},
\end{equation}

\noindent being $T_m(r)$ the weight of each topological charge $m$, given by \cite{24}:

\begin{equation}
\label{eq:04}
T_m(r) = \left\lbrace
\begin{array}{ll}
e^{-j\frac{m}{\ell}\frac{kr^2}{2f_{FR}}}e^{-j\frac{\pi m}{\ell N}}\mathrm{sinc}\left(\frac{m}{\ell N}\right)&\hspace{0.1cm}\textup{, if} \hspace{0.3cm} \ell-m=-\ell Nt,\hspace{0.1cm} \textup{$t=0,\pm 1, \pm 2,...$}\\
0&\hspace{0.1cm}\textup{, in other case} \hspace{0.3cm}
\end{array}
\right.
\end{equation}

\noindent with $\mathrm{sinc}(x)=\sin(\pi x)/\pi x$. From equation (\ref{eq:04}) it can be seen that $T_m$ is nonzero only when the equality $m=\ell(Nt+1)$ is satisfied. This shows how from a principal topological charge $\ell$ secondary topological charges $m$ can also be created depending on $N$ and $t$ (Fig. \ref{f:2} illustrates such dependence). Note that, for $t=0$ the principal topological charge $\ell$ is obtained. Because of the quadratic phase term, each secondary vortex will be focalized in a different plane depending on the value $\ell f_{FR}/m$.

Getting back on track with the field, its expression just after the off-axis Gaussian beam impinges the DVPL is

\begin{equation}
\label{eq:07}
U_1(r,\theta)=\sum_{t=-\infty}^{\infty}e^{-\frac{j\pi \left(Nt+1\right)}{N}}\mathrm{sinc}\left(\frac{Nt+1}{N}\right)e^{-\frac{jk\left(Nt+1\right)r^2}{2f_{FR}}}e^{j\ell\left(Nt+1\right)\theta}e^{-\frac{|\vec{r}-\vec{r}_{off}|^2}{w^2}},
\end{equation}

\noindent being $w$, $\vec{r}_{off}$ and $\vec{r}$, the beam radius, the position vector and the off-axis position vector of the Gaussian beam, respectively. Note that summation is done explicitly on $t$. After propagating the field a distance $Z$ from the lens (observation plane) it can be written as

\begin{eqnarray}
\label{eq:field prop}
U_2(\rho,\beta)&=&\frac{e^{jk(f+Z)}}{j\lambda f}\int_0^\infty\int_0^{2\pi} U_1(r,\theta)e^{\frac{jkr^2}{2f}(1-\frac{Z}{f})}e^{-\frac{jkr\rho}{f}\cos(\theta-\beta)}rdrd\theta\nonumber\\
&=&\sum_{t=-\infty}^{\infty}C_{Nt}\frac{e^{jk(f+Z)}}{j\lambda f} e^{-\frac{r_{off}^2}{w^2}}\times \int\limits^{+\infty}_{0} rdr\int\limits^{2\pi}_{0}d\theta\,\, e^{-r^2b_t}\,\,
e^{j\ell\left(Nt+1\right)\theta}\nonumber\\
&\times& \exp\left[\frac{2rr_{off}\cos(\theta+\theta_{off})}{w^2}\right]\exp\left[-\frac{jkr\rho}{f}\cos\left(\theta-\beta\right)\right],
\end{eqnarray}

\noindent where $f$ is the focal distance of the lens, $b_t=\frac{1}{w^2}-j\frac{k}{2}\left(\frac{1}{f}-\frac{Z}{f^2}-\frac{Nt+1}{f_{FR}}\right)$ and
\begin{eqnarray}
\label{eq:Cnt}
C_{Nt}=\exp\left[-\frac{j\pi \left(Nt+1\right)}{N}\right]\mathrm{sinc}\left(\frac{Nt+1}{N}\right).
\end{eqnarray}

Note that, from Eq. (\ref{eq:field prop}), if $\mathrm{Im}\{b_t\}= \frac{1}{f}-\frac{Z}{f^2}-\frac{Nt+1}{f_{FR}}=0$ a focal plane appears at:

\begin{equation}
\label{eq:11}
Z_t=f-\frac{\left(Nt+1\right)f^2}{f_{FR}}.
\end{equation}

Equation (\ref{eq:11}) indicates the positions on the optical axis where different optical vortices are formed according to the values of $N$ and $t$. The principal vortex appears at  $Z=f-f^2/f_{FR}$, corresponding to $t=0$, whereas for $t\neq0$ the positions of the SOVs are predicted.

Returning to Eq. (\ref{eq:field prop}), and because each term of the summatory can be treated as the field produced by a continuous SPP of topological charge $m$, by grouping the arguments of the last two exponentials and applying a definition of variables similar to that one reported in \cite{19},

\begin{eqnarray}
\label{eq:011}
\rho\cos\left(\beta\right)+\frac{jf}{kw^2}2r_{off}\cos\left(\theta_{off}\right)&=&\gamma\cos\left(\psi\right),\\[.3cm]
\rho\sin\left(\beta\right)-\frac{jf}{kw^2}2r_{off}\sin\left(\theta_{off}\right)&=&\gamma\sin\left(\psi\right),\\ \nonumber
\end{eqnarray}

\noindent with $\gamma$ and $\psi$ given by

\begin{eqnarray}
\label{condi}
\gamma^2&=&\rho^2-\frac{4f^2r_{off}^2}{k^2w^4}+j\frac{4f\rho r_{off}}{kw^2}\cos\left(\beta+\theta_{off}\right),\\[.3cm]
\tan\left(\psi\right)&=&\frac{\rho\sin\left(\beta\right)-j\frac{f}{kw^2}2r_{off}\sin\left(\theta_{off}\right)}{\rho\cos\left(\beta\right)+j\frac{f}{kw^2}2r_{off}\cos\left(\theta_{off}\right)},
\end{eqnarray}

\noindent the integral in Eq. (\ref{eq:field prop}) can be simplified to

\begin{eqnarray}
\label{12}
U_2\left(\rho, \beta\right)&=&\sum_{t=-\infty}^{\infty}C_{Nt}\frac{e^{jk\left(f+Z\right)}}{j\lambda f}e^{-\frac{r_{off}^2}{w^2}}\nonumber\\[.3cm]
&&\times \int_0^{+\infty}e^{-b_tr^2}\left[\int_0^{2\pi}e^{j\ell\left(Nt+1\right)\theta}e^{-\frac{jkr\gamma}{f}\cos\left(\theta-\psi\right)}d\theta\right]rdr.
\end{eqnarray}

\noindent Solving the integrals it is found that: 

\begin{eqnarray}
\label{eq:09}
U_2\left(\rho,\beta\right)&=&e^{jk(f+Z)}\sum_{t=-\infty}^{\infty}C_{Nt}j^{- \left((Nt+1)\ell+1\right)}e^{j\ell(Nt+1)\psi}e^{-\frac{r_{off}^2}{w^2}} \nonumber\\
&& \times\frac{k\sqrt{\pi}} {4fb_t^{3/2}}e^{-\frac{\eta^2}{2b_t}}\eta\left[I_{\frac{\left(|\ell\left(Nt+1\right)|-1\right)}{2}}\left(\frac{\eta^2}{2b_t}\right)-I_{\frac{\left(|\ell\left(Nt+1\right)|+1\right)}{2}}\left(\frac{\eta^2}{2b_t}\right)\right],\;\;\;\;
\end{eqnarray}

\noindent where $I_n(x)$ is the modified Bessel function and $\eta=\frac{k\gamma}{2f}$. Note that, $\eta$ can take complex values when $r_{off}\neq0$. Eq. (\ref{eq:09}) is a general form of the field at the observation plane since it considers both, an off-axis displacement of the Gaussian beam and the discretization of the phase mask. As we will see in the next sections, by analyzing this expression it is possible to observe effects on irradiance and phase that could not be predicted by previous works. Meanwhile, note that, if we take $r_{off}=0$ then $\gamma^2=\rho^2$, $\psi=\beta$  and Eq. (\ref{eq:09}) is reduced to the particular expression obtained in \cite{24} using a centered  DVPL. Additionally, if we take $N\longrightarrow\infty$,  a particular solution analogous to that one shown in \cite{19} using a CSPP can be obtained. 

\subsection{Field Analysis}

\begin{figure}[t]
\centering\includegraphics[width=8 cm]{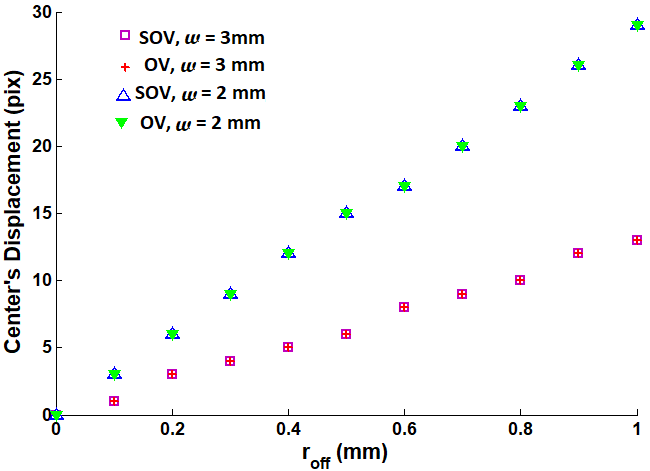}
\vspace{-0.3cm}
\caption{Vortex position at observation plane for different Gaussian beams of waist $w$ as a function of $r_{off}$. The simulation parameters are $N=2$, $f=200$ mm, $f_{FR}=-1.0$ m. For the principal vortex (OV) $\ell = -1$, $t=0$, and for the secondary vortex (SOV) $m=-1$, $t=-1$.}
\label{f:03}
\end{figure}

Now let's look at the effect of some terms of this equation. The complex nature of $\eta^2$ and $\psi$ (when $r_{off}$ is different from zero) is responsible for the asymmetry on the irradiance of the OVs. From Eq. (\ref{condi}) we know that the principal vortex is formed at position $(\rho,\beta)=(\frac{2r_{off}f}{kw^2},\theta_{off}-\frac{\pi}{2})$, similar to that shown in \cite{19}. Note that, the vortex position is independent of $N, \ell, t$ and $f_{FR}$, which are the parameters of the DVPL, but depends on $w$ and $f$, which corresponds to the Gaussian beam and optical setup parameters. So, the smaller the Gaussian beam radius or the larger is $f$, the greater off-axis displacement of the principal vortex singularity at the observation plane. In Fig (\ref{f:03}) we show an example for a principal charge $\ell=1$ and a secondary charge $m=-1$, for two different beam radius $w$. Note that there is no difference between the principal and secondary vortex displacements. The vortex position was obtained by following the centroid of the minimum of irradiance at the center of the ring-shaped pattern. The centroid was found using MATLAB \textit{regionprops} function, setting the \textit{centroid} property and a threshold corresponding to 0.2 the normalized irradiance.

\begin{figure}[t]
\centering
\includegraphics[width=8 cm]{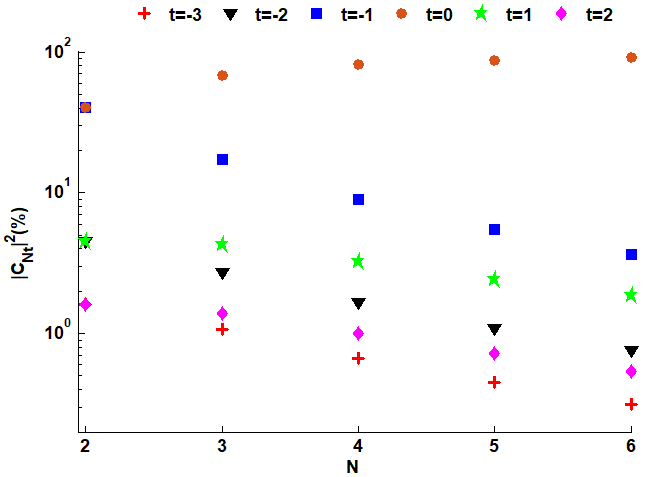}
\vspace{-0.3cm}
\caption{Overall energy (percentage) corresponding to each term of Eq. (\ref{eq:09}), $t=0$ for the principal vortex and $t=-3, -2, -1, 1, 2$ for the secondary vortices. For a better visualization, logarithmic scale was used. }\label{f:02}
\end{figure}

Continuing analizing Eq. (\ref{eq:09}), recall that $C_{Nt}$ contains the term $\mathrm{sinc}\left(\frac{Nt+1}{N}\right)$ (Eq. (\ref{eq:Cnt})) which controls the overall amount of energy corresponding to each vortex $m$ (or order $t$) in the expansion. In Figure \ref{f:02}, it is shown (in logarithmic scale) its weight corresponding to the principal vortex (order $t=0$) and to some secondary vortices (orders $t=-3,-2,-1,1,2$), for different $N$ discretization levels of the phase mask (notice that this is valid for any principal vortex $\ell$).
It can be seen that as discretization levels increase, the amount of energy available for secondary vortices decreases. For example, for the case of two levels, 40\% of the energy goes to the principal vortex, another 40\% goes to the first secondary vortex and the remaining 20\% is available for the remaining SOVs.

Recalling now on the quadratic complex term $\exp\left(-\frac{\eta^2}{2b_t}\right)$, it is directly related to the focusing of each vortex at a distance $Z_t$ when the imaginary part of $b_t$ becomes zero, allowing then the generation of a multifocal arrangement of optical vortices. To see this we take the real part of the complex-term argument and rearrange terms to obtain:

\begin{equation}
\label{eq:RePhaseTerm}
\exp\bigg(-\frac{\big(k^2/2f^2w^2\big)\big(\rho^2 - 4f^2r_{off}^2/k^2w^4 + 2f\rho r_{off}\cos\beta\big(1/f - Z/f^2 -(Nt+1)/f_{FR}\big)\big)}{4/w^4 + k^2(1/f -Z_t/f^2 - (Nt+1)/f_{FR})^2}\bigg)
\end{equation}

Because we are only interested to account for the optical field at the different focal planes, using Eq. (\ref{eq:11}) and defining $t$ as the order of interest to be seen at distance $Z_t$, and $t'$ for the vortices that are not focalized at distance $Z_t$, Eq. (\ref{eq:RePhaseTerm}) simplify to the following expression:

\begin{equation}
\label{eq:RePhaseTerm2}
\exp\Bigg(-\bigg(\frac{k^2}{2f^2w^2}\bigg)\frac{\rho^2 - \big(2fr_{off}\big/kw^2\big)^2 + 2f\rho r_{off}\cos\beta N(t-t')\big/f_{FR}}{4\big/w^4 + \big(kN(t-t')\big/f_{FR}\big)^2}\Bigg)
\end{equation}

Apart from a constant factor that is equal for all terms in the expansion, the numerator has three terms: the first one corresponds to a Gaussian beam centered at the optical axis, the second one corresponds to a shift of the Gaussian beam as a function of the displacement $r_{off}$ in accordance with the position of the singularity, and the third term corresponds to a radial deformation of the Gaussian beam shape as a function of the angle $\beta$ (Note that for the principal vortex, $t=t'$, this term is zero). More important for the distribution of the energy, is the denominator which is proportional to the fourth power of the inverse of the Gaussian beam radius. It reaches its minimum value $4/w^4$ for the vortex that focalizes at $Z_t$, and increases in multiples of $kN\big/f_{FR}$ depending on the order of the corresponding vortex with respect to the focalized vortex, i.e. increases for the non-focalized vortices.

\begin{figure}[t]
\centering
\includegraphics[width=8cm]{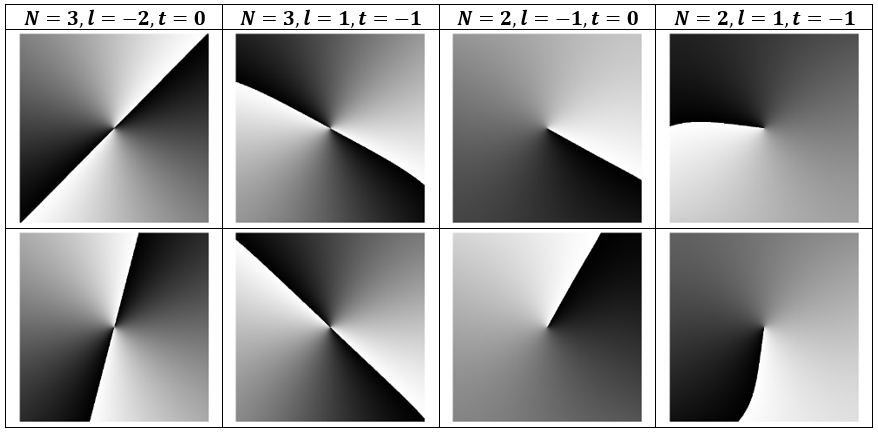}
\vspace{-0.3cm}
\caption{Phase rotation with the additional factor $j^{-Nt}e^{-\frac{j\pi(Nt+1)}{N}}$ term, with $r_{off}=0$. The first row representing the phase distribution when we take the total field and the second row representing the phase response when this factor is neglected. }\label{f:4}
\end{figure}

Finally, the factor $j^{-Nt}e^{-\frac{j\pi(Nt+1)}{N}}$ is be responsible of a constant phase rotation presented in Fig. \ref{f:4}.

\section{Signal to noise ratio of secondary vortices}
\label{S:SNR}

We now want to know how much does a non-focalized vortex affect the vortex of interest, of order $t$, at plane $Z_t$. To do this we compare at plane $Z_t$ the optical power of the vortex of order $t$ with respect to the optical power of the rest of the vortices (noise), measuring the signal-to-noise ratio (SNR)

\begin{equation}
\label{eq:SNR}
\mathrm{SNR}_{dB} = 10\log_{10}\Bigg( \frac{P_t}{P_{noise}}\Bigg)
\end{equation}

\noindent $P_t$ is obtained by integrating the square module of the term $t$ of interest in Eq. (\ref{eq:09}) over the area of influence, which is defined as the area subtended by a radius from the center of the vortex to the distance where the maxima of the vortex reduces to a half. $P_{noise}$ is obtained by integrating the other terms of the summation in Eq. (\ref{eq:09}) over the same area, and summing. We then analyze the effects of using different parameters in the system. Without loss of generality, off-axis displacements in the $x$-axis direction ($\theta_{off}=0$) are only considered.

\begin{figure}[t]
\centering
\includegraphics[width=14 cm]{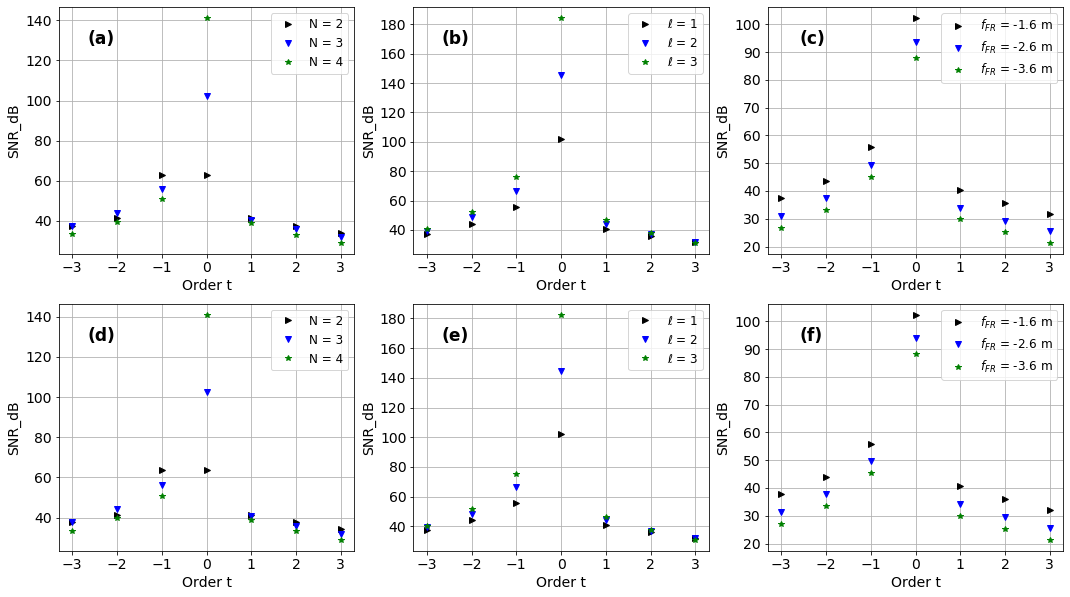}
\vspace{-0.5cm}
\caption{SNR as function of orders for the on-axis condition (first row) and the off-axis one (second row). All cases with $f = 200$ mm, $\lambda = 532$ nm, $w = 6.0$ mm, being (a) $\ell = 1$, $f_{FR} = -1.6$ m and variable $N$; (b) $N = 3$, $f_{FR} = -1.6$ m and variable $\ell$; and (c) $\ell = 1$, $N = 3$ and variable $f_{FR}$.}\label{f:SNR-on}
\end{figure}

Figure \ref{f:SNR-on} shows the SNR when the beam is on-axis ($r_{off}=0.0$ mm, first row) and off-axis ($r_{off}=1.0$ mm, second row). In all cases $w = 6$ mm, $f = 200$ mm, $\lambda = 532$ nm, and the sum in Eq. (\ref{eq:09}) is done only with orders $t-10 \leq t \leq t+10$. Sub-figures \ref{f:SNR-on}(a) and (d) depict the SNR for different phase-discretization levels, whereas Subfigs. \ref{f:SNR-on}(b) and (e) refer to different topological charges, and Subfigs. \ref{f:SNR-on}(c) and (f) present different $f_{FR}$ values. As a first point to highligth is that there is not any apparent difference between the SNR for the on-axis and off-axis cases. Moreover, from the figures it is apparent that for all cases the SNR is greater than 20 dB, meaning that any of the secondary vortices could be employed, for example, in a metrological application. In all cases (with the exception of $N=2$) the order zero has a better SNR than other orders. Besides, other characteristics can be highlighted. Figures \ref{f:SNR-on}(a) and (d), and figs. \ref{f:SNR-on}(b) and (e) present a similar behavior: a marked variation in the SNR for $t=0$, a small decrease for $t=-1$ when $N$ increases ($\ell$ decreases) and almost no variation for the other orders. On the other hand, Figs. \ref{f:SNR-on}(c) and (f) present an almost constant variation of the SNR for all orders. The reduction in the SNR when $f_{FR}$ gets larger is because the DVPL tends to become a discrete SPP.



Having analyzed the SNR in several conditions and found that in all cases there is not considerable degradation of OVs being on or off axis, we now focus our attention to the phase and irradiance distribution of this multifocal arrangement of asymmetric optical vortices at each focal plane.

\section{Analysis of irradiance and phase on secondary asymmetric optical vortices}
\label{S:3}

In this section, principal vortices with topological charges $l=-3,-2,-1$ and secondary OVs with the same topological charges, are analysed. As in the previous section, off-axis displacements in the $x$-axis direction ($\theta_{off}=0$) are only considered. The optical setup parameters are $w=3.0$ mm, $\lambda=532$ nm, $f=200$ mm and $f_{FR}=-1000$ mm, for all simulations. It should be noted that, taking into account the overall energy distribution given by Eq. (\ref{eq:04}), we only consider the first main orders ($-3\leq t\leq 2$) which contribute the most to the vortex formation. Additional terms of Eq. (\ref{eq:09}) are disregarded.

\begin{figure}[t]
\centering\includegraphics[width=14 cm]{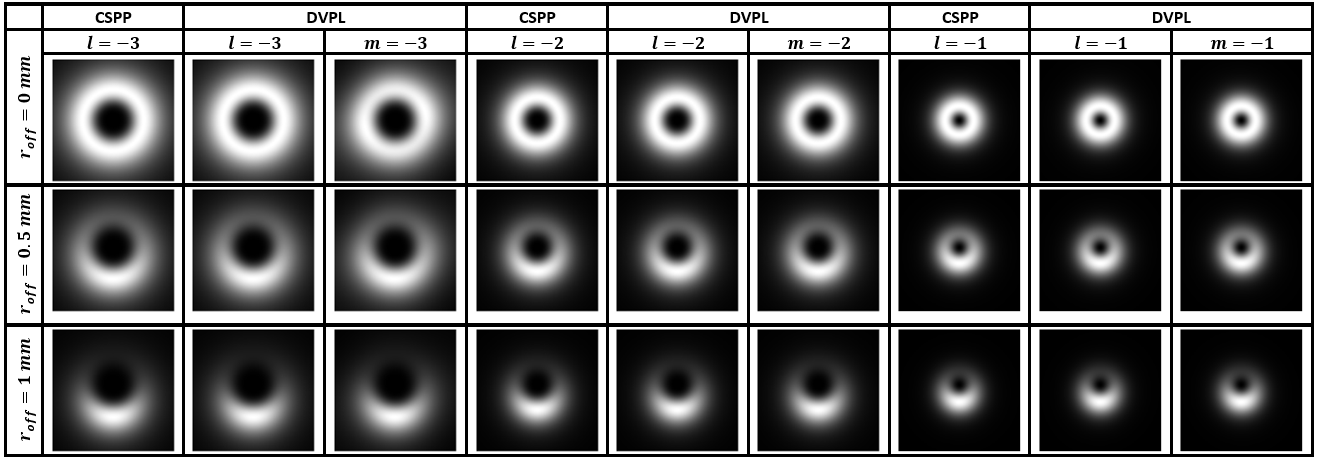}
\vspace{-0.5cm}
\caption{Comparison of an OV irradiance obtained using a CSPP, a DVPL principal OV, and a DVPL SOV, for different off-axis displacements, as indicated in the figure. To obtain $m = -3$ we used $\ell=1$, $N=2$, and $t = -2$; for $m=-2$ we used $\ell = 2$, $N = 2$, and $t = -1$; and for $m=-1$ we used $\ell = 1$,  $N=2$ and $t = -1$. For the principal vortices the respective topological charge values are directly programmed on the mask using $N=2$ (the area observed is 1 mm x 1 mm). The irradiance is normalized to facilitate the visualization.}
\label{f:3}
\end{figure}

\subsection{Irradiance distribution}
In this subsection, we compare the irradiance pattern between a CSPP, a DVPL principal vortex, and a DVPL secondary vortex (all with the same charge), for different off-axis displacements. In Fig. \ref{f:3} we present the corresponding irradiances for topological charges -1, -2, and -3, and for $r_{off}=$ 0.0 mm, 0.5 mm, and 1.0 mm. Irradiance patterns are normalized to facilitate comparison. This is justified because a high SNR was obtained in the previous section. 

\begin{figure}[t]
\centering\includegraphics[width=8 cm]{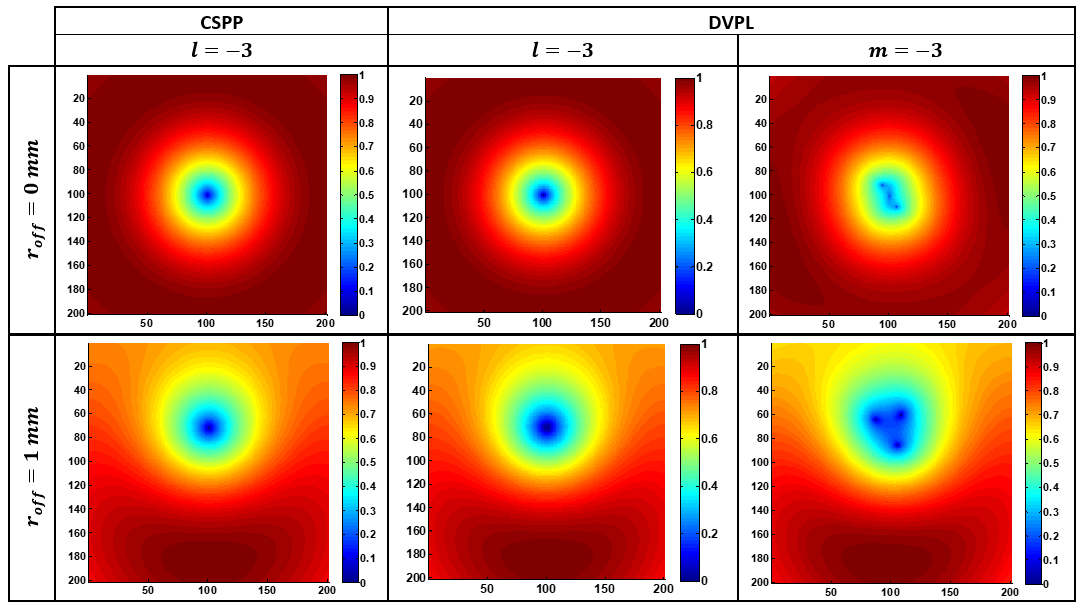}
\vspace{-0.4cm}
\caption{ Close view of the irradiance distribution emphazising the ubication of the singularity positions for  principal vortices generated with a CSPP (first column), a DVPL (second column) being $l=-3$ and a secondary vortex generated with a DVPL (third column) with topological charge $m=-3$, with the beam centered (row 1) and impinging 1 mm off-axis (row 2) (the area observed is 0.56 mm x 0.56 mm). The same parameters as in Fig. \ref{f:3} are employed.}
\label{f:5}
\end{figure}

When comparing the different patterns no appreciable difference is observed. In fact, if the method proposed by Anzolin et al. \cite{19} is applied the results will be the same irrespective the source of the OV and will only depend on the topological charge value. This explains why the experimental results in Ref. \cite{22}, for an off-axis Gaussian beam impinging onto a binary VPL, agrees with the results presented by Anzolin et al. 

\begin{figure}[!thb]
\centering\includegraphics[width=14 cm]{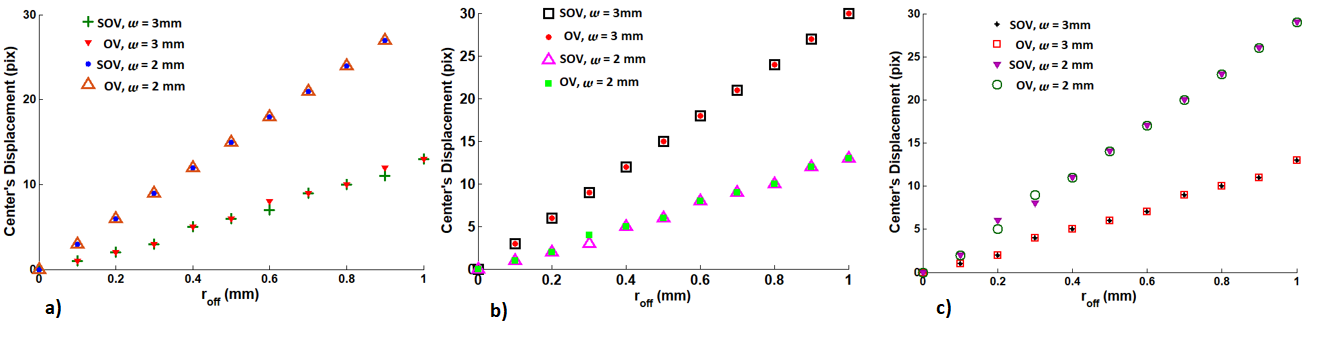}
\vspace{-0.6cm}
\caption{Centroid position at observation plane for different Gaussian beams of waist $w$ as a function of $r_{off}$ with  $f=200$ mm and $f_{FR}=-1000$ mm. The corresponding principal and secondary vortices have the same topological charge. The parameters are: a) $\ell = -3$ and $m = -3$, b) $\ell = -2$ and $m = -2$, and c) $\ell = -1$ and $m = -1$.}
\label{fi:03}
\end{figure}

However, a closer look into the singularity shows a difference in the case of the SOV due to the effect of the other SOVs and principal vortex. In Fig. \ref{f:5} is shown the case for a topological charge -3 and an off-axis displacement of 0 mm and 1 mm. For the CSPP and the principal topological charge of the DVPL, the patterns are almost identical and in each case, only one singularity is recognized. On the contrary, for the secondary topological charge of the DVPL, the separation of the singularities is evident. However, it is worth noting that the changes are located around a centroid that experiences the same displacement of the singularities of the first two cases (CSPP and DVPL principal vortex), predicted by Eq. (\ref{condi}). The centroid corresponds to the position expected for the principal vortex in the case of a DVPL, or the vortex in a CSPP. In Fig. \ref{fi:03} the centroid position as a function of the off-axis displacement is shown for topological charges $m=$ -3, -2, and -1, obtained from DVPL simulations. Like in Fig. \ref{f:03} the displacement is determined following the centroid of the irradiance minima. As was previously found in the analysis related to Eq. (\ref{condi}) and depicted in Fig. \ref{fi:03}, the beam radius $w$ directly affects the displacement of the centroid for a given off-axis displacement of the beam: the greater the radius, the smaller the centroid's displacement. From the figures, it is also verified that the centroid's position does not depend on the topological charge. Finally, it is shown that the centroid position for the SOV coincide with the principal OVs for the topological charges and the off-axis displacements considered.

\begin{figure}[tb]
\centering\includegraphics[width=14 cm]{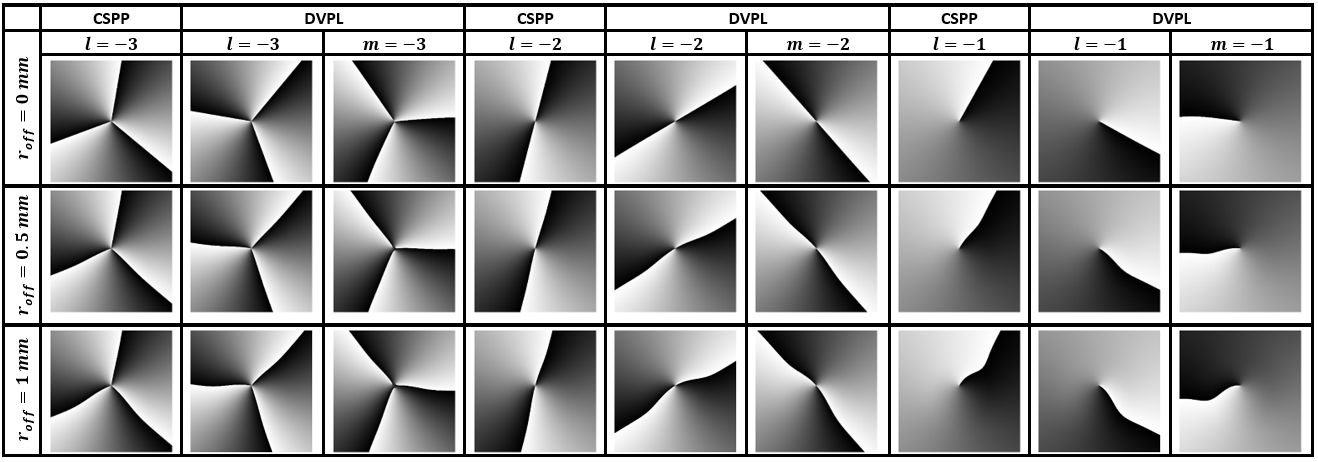}
\vspace{-0.6cm}
\caption{Comparison of an OV phase obtained using a CSPP, DVPL principal OV, and a DVPL SOV, for different off-axis displacements, as indicated in the figure. A gray-scale is used where the phase run continously from $0$ (black) to $2\pi$ (white). The same parameters as in Fig. \ref{f:3} are employed.}
\label{f:7}
\end{figure}
 
  \begin{figure}[htb]
\centering\includegraphics[width=8 cm]{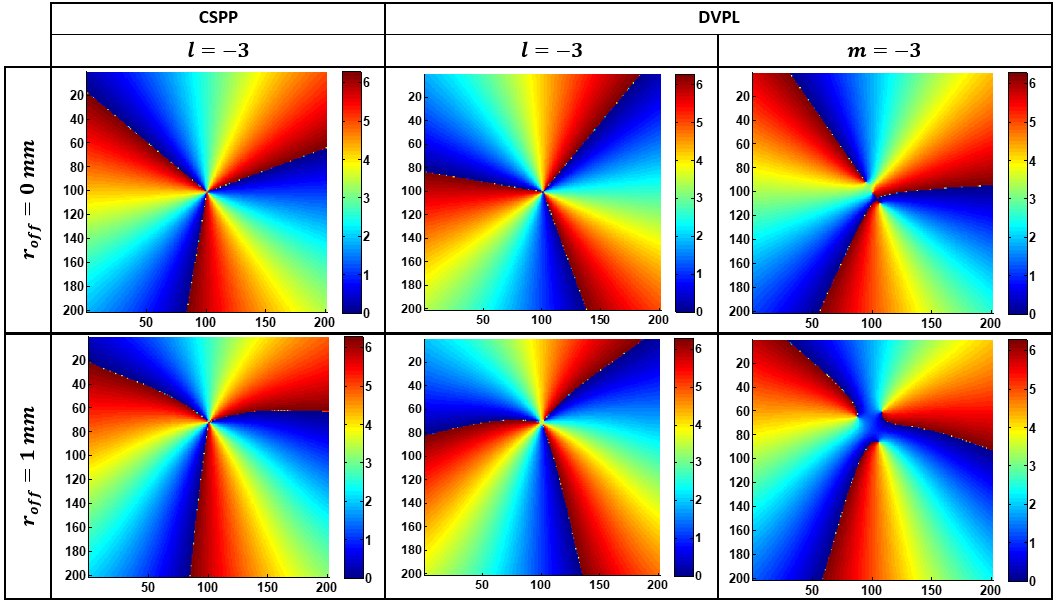}
\vspace{-0.3cm}
\caption{Close view of the phase distribution stressing the ubication of the singularity positions of principal vortices generated with a CSPP (first column), a DVPL (second column) being $l=-3$ and a secondary vortex generated with a DVPL (third column) with topological charge $m=-3$, with the beam centered (row 1) and impinging 1 mm off-axis (row 2). The same parameters as in Fig. \ref{f:5} are employed.}
\label{f:9}
\end{figure}

\subsection{Phase response}
In Fig. \ref{f:7} and Fig. \ref{f:9} the phases corresponding to the irradiance patterns of Fig. \ref{f:3} and Fig. \ref{f:5} are shown, respectively. From the figure, three main characteristics can be highlighted: phase singularity (or the centroid for the SOV case) are displaced as expected, the discontinuity lines of the phase show a curvature proportional to the displacement, and the angles between the discontinuity lines are also modified. Besides, for the SOVs, as in the irradiance case, the separation of the phase singularities change proportional to the displacement. In Fig \ref{f:9}, the phase distribution corresponding to topological charge -3 is shown to appreciate the changes and differences in more detail.

\section{Phase features as displacement indicator}

\begin{figure}[htb]
\centering\includegraphics[width=12 cm ]{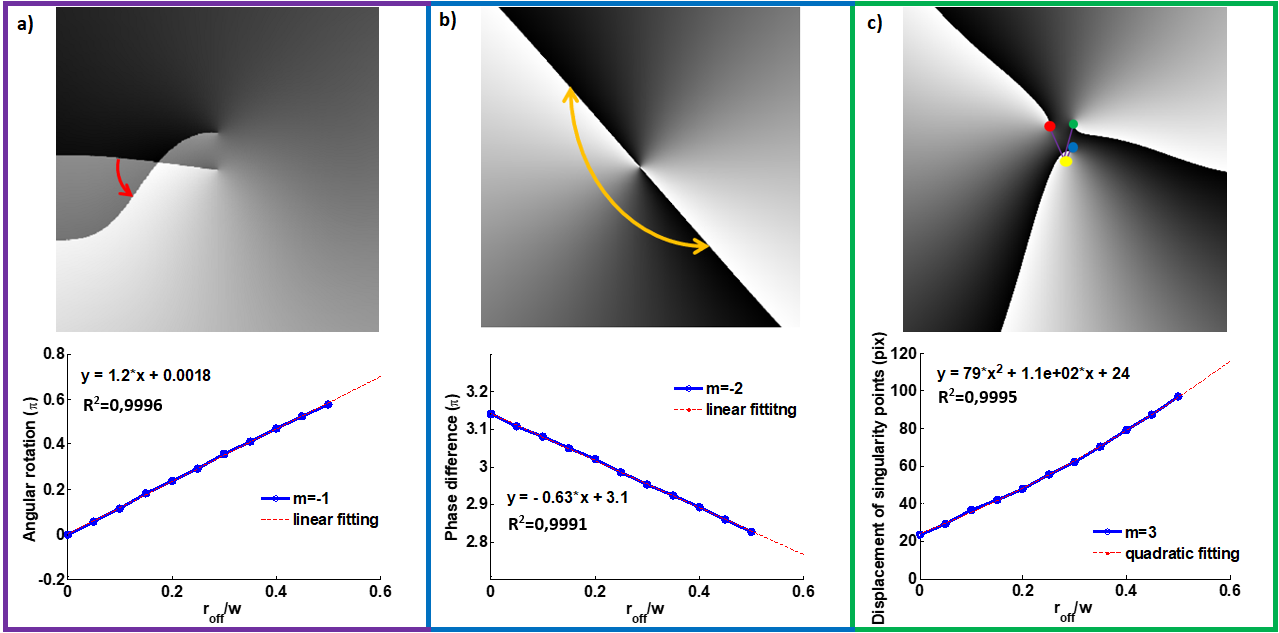}
\caption{Traslation metrics using the phase distribution of secondary optical vortices with topological charges: (a) $m=-1$, (b) $-2$, and (c) $m=-3$.}
\label{f:8}
\end{figure}

The different behavior of the phase singularity position and dislocation lines could be used as features to follow the off-displacement of the Gaussian beam. With this in mind, we have proposed three different metrics for three different topological charges to study its response to displacements. In Fig. \ref{f:8} (first row) we show the principles of the metrics for topological charges -1, -2 and -3.

Let us start with the phase distribution for $m=-1$, Fig. \ref{f:8}a (top). The main effect in the OVs with this charge is an angular rotation of its discontinuity line. With this in mind, the angular rotation as a function of off-axis displacement of the Gaussian beam is analyzed. As a result, a linear dependence is found as depicted in Fig. \ref{f:8}a (bottom). Regarding the OVs with charge $m=-2$, the angle between the discontinuity lines accounts, see Fig. \ref{f:8}b (top). For $r_{off}=0$ this separation is equal to $\pi$, but as long as the off-axis displacement increases the lines of discontinuity approach each other. If the phase difference between discontinuity lines is plotted as a function of the off-axis displacement, a linear dependence is observed, as can be seen in Fig. \ref{f:8}b(bottom). Finally, for $m=-3$ (Fig. \ref{f:8}c (top)), as an indicator for the displacement, the sum of the distances of each singularity point (red, green and blue point) with respect to the centroid (yellow point) is evaluated. In this case, as is shown in Fig. \ref{f:8}c (bottom),  a quadratic dependence is found. For $r_{off}=0$ the minimum distance obtained is different of zero since for SOVs, even with the Gaussian beam centered, there is a slight separation of the singularity caused by the superposition of other defocused vortices that arrive at that plane of observation. These changes in the phase as a function of the off-axis displacement of SOVs open new possibilities to establish other measurement indicators.
 
 \section{Conclusions}
\label{S:5}
In this work, an analytical expression of the optical field produced when an off-axis Gaussian beam is diffracted by a discrete vortex producing lens, giving a multifocal arrangement of secondary optical vortices, is found. The analytical expression is more general than previous ones since it allows us to observe behaviors in the phase and irradiance distribution of the vortices that had not been previously reported.
The contribution of each term in the optical field is analyzed, showing that the results are dependent on system parameters such as discretization levels, Gaussian beam size, and the order $t$. Regarding the intensity, as a result of the discretization of the VPL, it is observed for the SOV with $m=-3$ that the vortex is split into three unitary vortices. Besides, if the misalignment appears,  the separation of the phase singularities of the vortices increases. Nevertheless, the centroid for the separated phase singularities is always located in the position corresponding to the phase dislocation for the case of a CSPP. Focusing on the phase of the field, SOVs have notorious changes in their phase distributions. They tend to present greater rotations due to the misalignment, accompanied by a separation of discontinuity lines in the phase. Therefore, there is a displacement of the singularity of the phase associated with the misalignment and a separation of the lines of discontinuity associated with the DVPL. These behaviors give rise to the possibility of using the phase as a parameter to measure off-displacements. We have also shown that all the SOVs have a good SNR and no noise problems appear, being the only important parameter for detection the energy or sensitivity available for the particular SOV of interest.

\section*{Acknowledgment}

N. L. acknowledges the receipt of the grant from the Abdus Salam International Centre for Theoretical Physics, Trieste (Italy), Centro Latino-Americano de Física (Brazil) (this work is partially supported by the ICTP-CLAF agreement AF-13) and to  Colciencias Convocatoria 785 Doctorados Nacionales. E. R. thanks Universidad de Antioquia U de A for financial support. D.A. thanks financial support from ANCYT PICT-2015 3385 (Argentina). J.A.G acknowledges the support from Politécnico Colombiano Jaime Isaza Cadavid.


\section*{References}





\end{document}